\documentclass[%
 reprint,
 amsmath,amssymb,
 aps,
pra,
]{revtex4-2}

\usepackage{graphicx}
\usepackage{dcolumn}
\usepackage{bm}
\usepackage{caption}
\usepackage{ragged2e}
 \usepackage[utf8]{inputenc}

 \usepackage[usenames,dvipsnames]{xcolor}

\begin{document} 
\title{Efficient, ever-ready quantum memory at room temperature for single photons} 
\author
{Anthony C. Leung,$^{1,2\dagger}$ W. Y. Sarah Lau,$^{1,2,3,4\dagger}$ Aaron D. Tranter,$^{1,2}$
Karun V. Paul,$^{1,2}$ Markus Rambach,$^{3,4}$ Ben C. Buchler,$^{1,2}$
Ping Koy Lam,$^{1,2}$ Andrew G. White,$^{3,4}$ Till J. Weinhold$^{3,4 *}$\\
\normalsize{$^{1}$ARC Centre of Excellence for Quantum Computation and Communication Technology}\\
\normalsize{$^{2}$Department of Quantum Science and Technology, Research School of Physics,}
\normalsize{The Australian National University, Acton, ACT 2601, Australia}\\
\normalsize{$^{3}$ARC Centre of Excellence for Engineered Quantum Systems}\\
\normalsize{$^{4}$School of Maths and Physics, University of Queensland, Brisbane, QLD 4072 Australia}\\
\normalsize{$^\dagger$These authors contributed equally to this work.}\\
\normalsize{$^\ast$Corresponding author; E-mail:  t.weinhold@uq.edu.au}
}
\date{\today}

\begin{abstract}
  Efficient quantum memories will be an essential building block of large scale networked quantum systems and provide a link between flying photonic qubits and atomic or quasi-atomic local quantum processors. To provide a path to scalability avoidance of bulky, difficult to maintain systems such as high vacuum and low temperature cryogenics is imperative. Memory efficiencies above 50\% are required to be operating above the quantum no-cloning limit. Such high efficiencies have only been achieved in systems with photon sources tailored to the memory bandwidth.  In this paper we explore the combination of an ultralow spectral bandwidth source of single photons from cavity-enhanced spontaneous parametric down-conversion with a gas-ensemble atomic memory.  Our rubidium vapour gradient echo memory achieves 84$\pm$3 \% recall efficiency of single photons: a record for an always-ready, hot, and vacuum system free optical memory.
\end{abstract}
\maketitle 

\section*{Introduction}

Harnessing the full alluring potential of quantum technologies will require linking quantum systems locally and globally \cite{kimble_internet,Wehner2018,Wei2022} to share resources and enhance capabilities: similar to the multi-processor CPUs of local computers, or the vast computing power that comes from connecting data centers to form the internet. Quantum interconnects will be an essential element in  linking the nodes of a quantum network consisting of sensors \cite{Gottesman2012}, distributed quantum computers  \cite{Jiang2007,Nickerson2014}, and quantum clock signals \cite{Komar2014}. To ensure information fidelity and practical transmission rates over long distances, the quantum network channels need to be optical \cite{Lee2019}, like their classical counterparts. Synchronisation between different systems will require the availability of quantum memories \cite{Heshami2016} to store and relay the qubits and to build quantum repeaters \cite{Briegel1998} that enable long distance communications.

A wide variety of quantum memory protocols have been demonstrated each with their own set of advantages and disadvantages in terms of performance metrics in efficiency, storage time, frequency bandwidth and other metrics.  Solid-state systems excel in storage times of the order of one minute \cite{Heinze_2013_PRL} and the potential for many hours \cite{Zhong_2015_Nat}.  Recall efficiencies around 90\%  were demonstrated using both warm \cite{Hosseini_Sparkes_Campbell_Lam_Buchler_2011} and laser-cooled gaseous alkali atoms \cite{Hsiao_2018_PRL,Cho_Campbell_Everett_Bernu_Higginbottom_Cao_Geng_Robins_Lam_Buchler_2016}. Memories using off-resonant Raman transitions have demonstrated THz bandwidths \cite{England_PRL_2015} and time-bandwidth products around 100 \cite{Guo_NatComm_2018}. 

Efficient storage of the photonic qubit in these atom-based memories requires matching memory and photon source features. Due to the complexity of engineering a quantum light source to optimally integrate with the quantum memory's temporal, spectral, and spatial acceptance bandwidth, most demonstrations have relied on attenuated classical light sources rather than true quantum light. But weak coherent states cannot replace single photons for applications such as photonic quantum computation \cite{Knill_Laflamme_Milburn_2001}, and physics at the quantum limit may exhibit fundamentally different interactions \cite{Fan_CrossKerrBreakdown_2013,Vural-18}. Hence operation with tailored quantum light sources is vital for real-world relevant benchmarking of quantum memories \cite{Tsai_2020_PRR,Heller_2021,Yu_2021}. 
To date the most efficient realisations employed electromagnetically induced transparency in a rubidium magneto-optical trap (MOT) for the memory and four-wave mixing or spontaneous Raman scattering in a separate cold atomic ensemble as the single photon source \cite{Wang_Li_Zhang_Su_Zhou_Liao_Du_Yan_Zhu_2019, Cao_Hoffet_Qiu_Sheremet_Laurat_2020}. These produced an outstanding recall efficiency of $85\%$, but the duty-cycle --- the fractional time period where the memory is ready for read/write operations--- was limited to $3\%$ by the loading and cooling steps of the MOTs \cite{Wang_Li_Zhang_Su_Zhou_Liao_Du_Yan_Zhu_2019}.

If quantum networks are to scale then highly efficient quantum memories with high duty-cycles are a necessity. Any efficiency below unity ultimately limits the range and depth of the network, adding to this low duty-cycles massively increases operational overheads.  Utilisation of memories based on hot or room-temperature atomic vapors can alleviate the loading and cooling steps, but typically suffer from added noise \cite{SFWMNoise,Michelberger_2015}, low storage times \cite{ORCA-PhysRevA.97.042316}, or both.

Here we report microsecond storage of single photons, with 
recall efficiencies up to 84 $\pm$ 3\%, in an always ready memory system. We achieve this by integrating a hot vapor rubidium gradient echo memory (GEM) \cite{Hosseini_Sparkes_Hetet_Longdell_Lam_Buchler_2009,Hosseini_Sparkes_Campbell_Lam_Buchler_2011, Hosseini_Campbell_Sparkes_Lam_Buchler_2011,Cho_Campbell_Everett_Bernu_Higginbottom_Cao_Geng_Robins_Lam_Buchler_2016} with a tailored single-photon source based on cavity-enhanced spontaneous parametric down-conversion (SPDC) \cite{Rambach_Nikolova_Weinhold_White_2016}. 
Our system is scalable: the first quantum memory and photon source combination that is vacuum-system free, non-cryogenic, free of an operational duty-cycle and surpasses the no-cloning \cite{Wootters1982} threshold.

\section*{Results}

\begin{figure}[t]
\vspace{0cm}
  \centering
  \includegraphics[width=.98\linewidth]{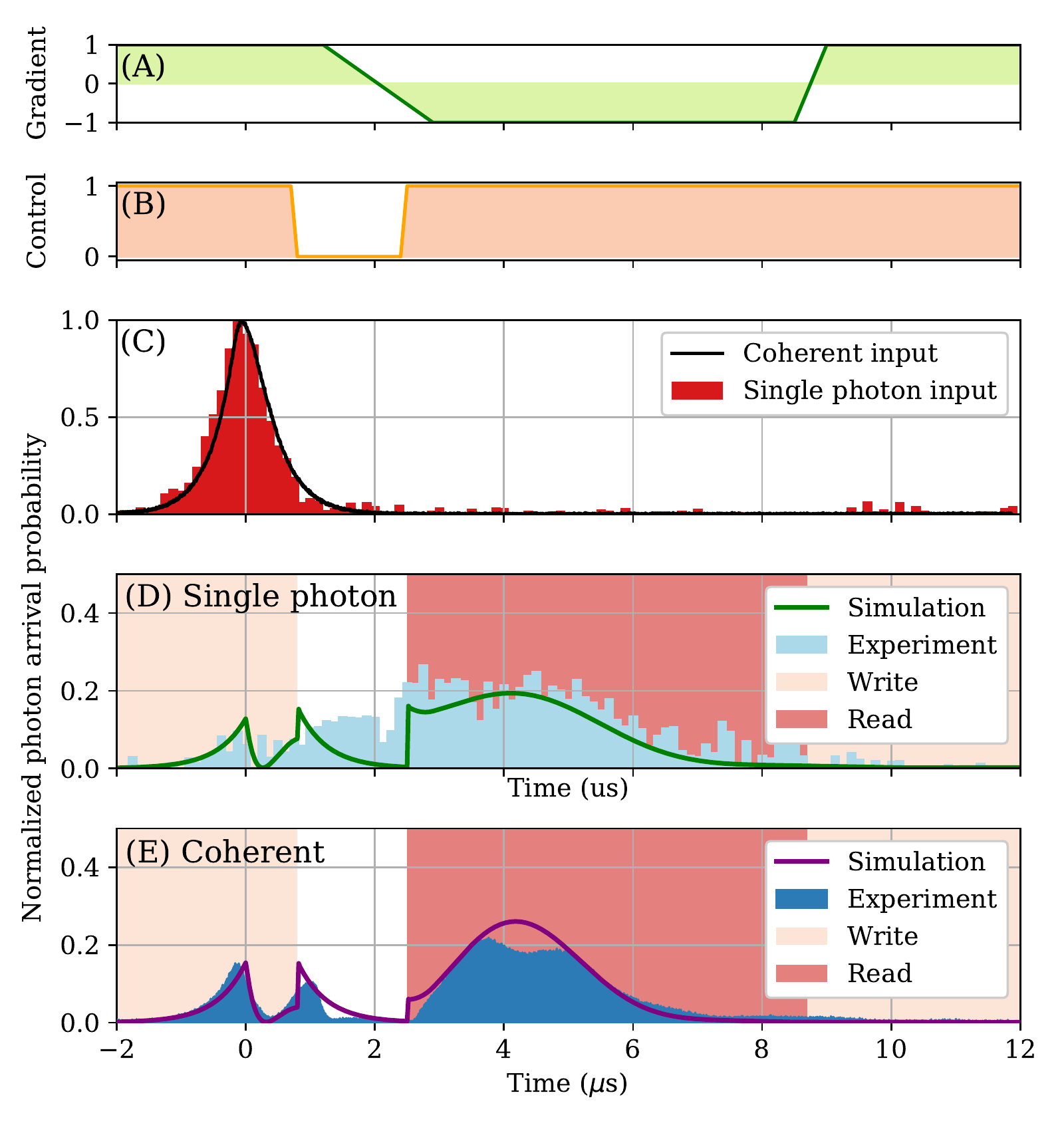}
  \vspace{-0.5cm}
  \captionof{figure}{\textbf{Memory operation and storage results}. The timing for \textbf{(A)} the magnetic field gradient, and \textbf{(B)} control field switching.  These determine the input phase (light pink), the storage phase (white) and the recall phase (dark pink) indicated in (D and E). \textbf{(C)} Experimentally collected coincidence data displayed using histograms of the input single photon state (red) overlaid with the corresponding input coherent state (black).  \textbf{(D)} Experimental coincidence data of single photon state stored and recalled after 4~$\mu$s (light blue) overlaid with simulated results (green) showing the sharp onset of the recall signal due to the larger spectral bandwidth of the single photons causing bandwidth mismatch with the memory.  \textbf{(E)}  Experimental data of corresponding coherent state storage (blue) overlaid with simulated results (purple) showing bandwidth matched storage.}
  \label{expvscovssim}
  \vspace{0cm}
\end{figure}

Typical storage and recall of single photons for a 4~$\mu$s duration is presented in Figure~\ref{expvscovssim}. Dark/light pink shaded regions indicate the times when the memory is in the read/write modes respectively as indicated by the magnetic field gradient state and presence of the control field shown in Figures~\ref{expvscovssim}A and \ref{expvscovssim}B respectively. The red histogram of Figure~\ref{expvscovssim}C indicates single photons arriving at the memory during the ``no memory'' stage while the control field is turned off: they are detected as coincidences with a herald detection on the idler photon produced by the down-conversion source (see Methods for details). The blue histogram in Figure~\ref{expvscovssim}D indicates photon coincidences with the memory active and set to delay the signal by 4~$\mu$s. 
We exploit the temporal correlation between the twin photons emitted from the SPDC source to provide a high quality single photon signal through the detection in the coincidence basis \cite{Hong,Pittman}. This powerful technique enables us to conclusively identify the single photon signal in front of a background of leaked control field photons of $\sim$ 5-10k counts/s (See Methods for details). Such strong background would have rendered other, more sensitive single photon characterisation measurements like cross and auto-correlation \cite{first-g2} fruitless. 

We compare the performance of the memory acting on single photons to the storage of a coherent state. We temporally shape attenuated pump laser light using an acousto-optic modulator (AOM) so that it possesses the same temporal profile as the single photons. The input pulse without storage is shown as the black solid line in Figure~\ref{expvscovssim}C, which agrees well with the measured not-stored single photons. Recall of the coherent state from memory is shown in Figure~\ref{expvscovssim}E: it is strikingly similar to the recalled photon profile. The main difference being how sharply the single photon recall appears as the control field was turned back on. This phenomena arises from different spectral bandwidths of the two states; single photons generated by SPDC have a wider spectral bandwidth than the coherent state generated directly by the narrow linewidth laser. This is elucidated from a simulation for the same input temporal profile but with different relative bandwidths shown in Figures~\ref{expvscovssim}D and \ref{expvscovssim}E.  The simulated single photon output was obtained by increasing the input to memory bandwidth ratio by 25\% from the simulated coherent output and caused the recall to appear earlier, and flattened as observed in the single photon experimental results. This slight spectral mismatch delivers a substantial difference in the recall efficiency: the coherent state recall reaches a maximal value of 55$\pm$1\% compared to the 84\% for the single photon.

\begin{figure}[t]
  \centering
  \vspace{-0.8cm}
  \includegraphics[width=1\linewidth]{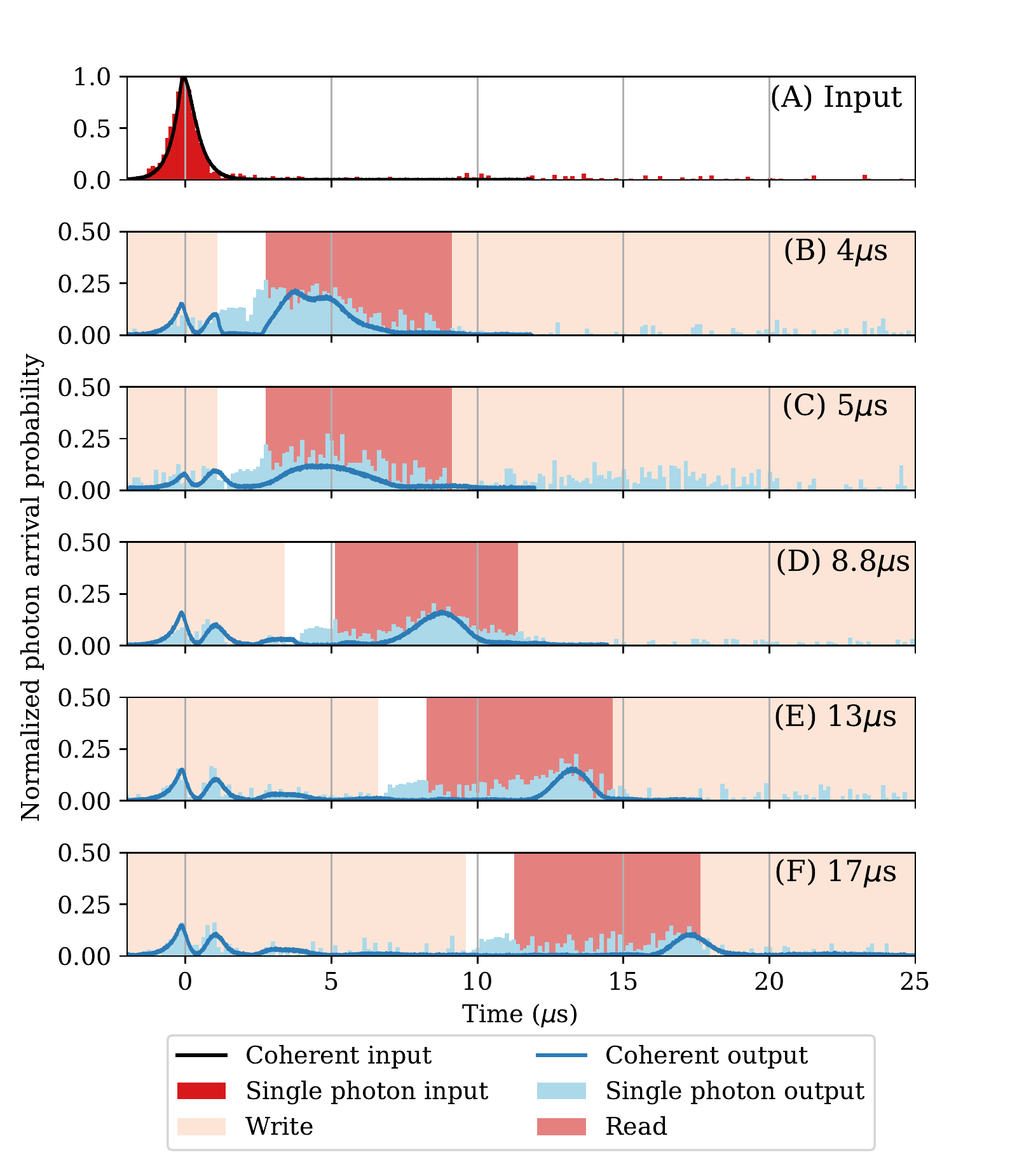}
  \captionof{figure}{\textbf{Single photon storage over a range of storage times.} Coincidence histograms demonstrating storage of single photons at a range of storage times along with their matching results from coherent state storage.}
  \label{lifetime}
\end{figure}

Storage and recall experiments were completed at varying storage times shown in Figure~\ref{lifetime}. We classified as recalled, any coincident photons detected after the control field was switched back on and before the magnetic field gradient was flipped back to write mode. The data clearly shows the centre of the recalled signal shifting with the extended storage time for both the coherent state (solid lines) and the single photons (red and light blue histograms), which retain their scattered signal preceding the main recall associated with the larger bandwidth.  

\begin{figure}[t]
\centering
    \includegraphics[width=0.9\linewidth]{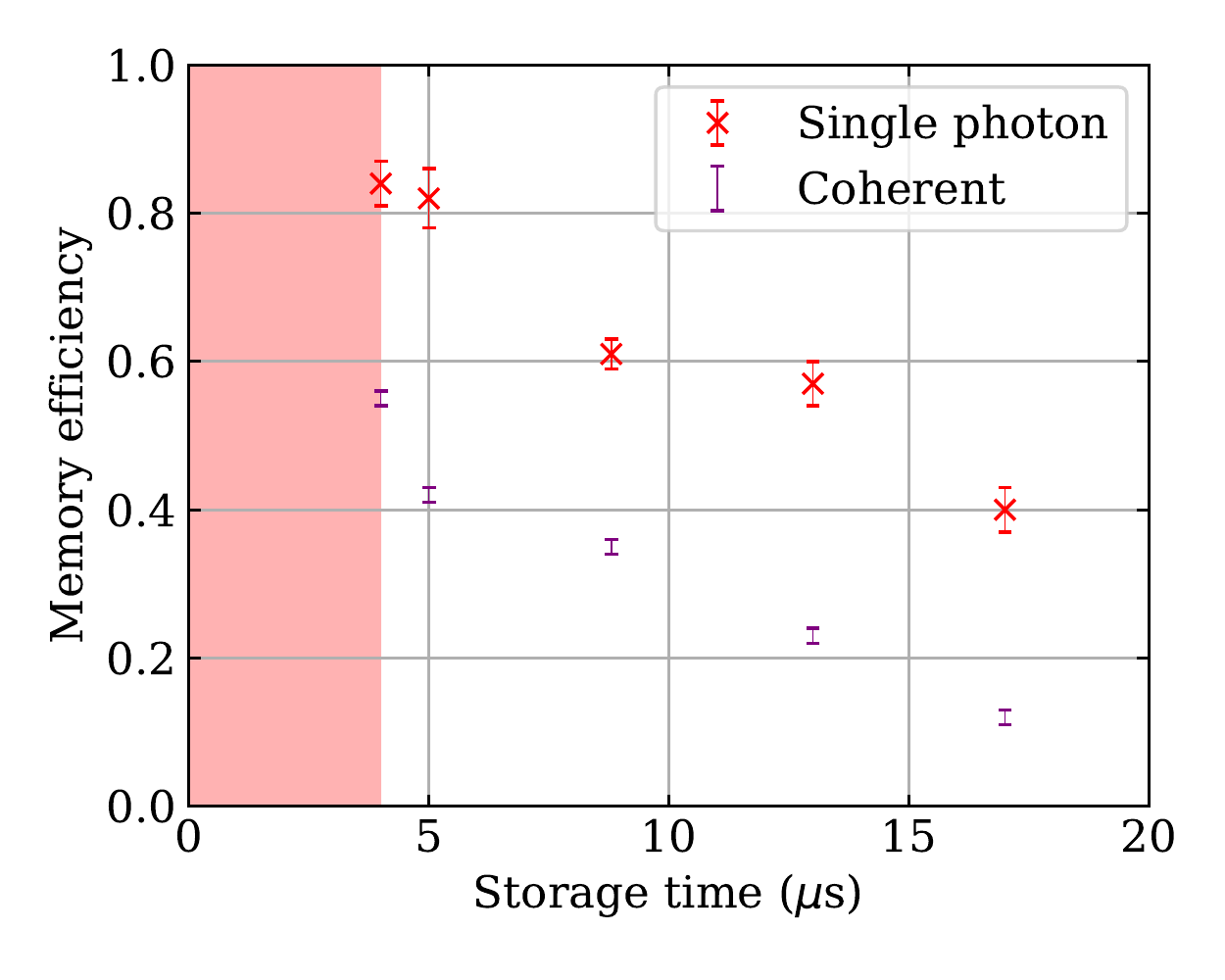}
    \captionof{figure}{\textbf{Memory recall efficiency at different storage times.}  The red shaded area shows storage times shorter than our minimum storage duration.  Memory efficiencies for the different times are shown for both the single photon states (Red markers) and the coherent state (Purple markers). }
    \label{effvstime}
\end{figure}

Recall efficiency decreases with an increasing storage time, in part due to addressed atoms moving out of the interaction region via atomic motion in the warm vapour and because of atom-atom collisions. The reduction in recall efficiency is clearly visible as reduced area of the recalled signal in Figure~\ref{lifetime} and is plotted in Figure~\ref{effvstime}. The efficiency drops from 84 $\pm$ 3\% at 4~$\mu s$ to 57 $\pm$ 3\% at 13~$\mu$s before falling below the no-cloning limit \cite{Wootters1982} at 17~$\mu$s when we recall only 40 $\pm$ 3\% of the stored signal. Storage times shorter than 4~$\mu s$ are inaccessible due to limitations of our setup. Error bars are derived from the propagated statistical noise in each measurement, directly related to the total number of coincidence counts measured. Our protocol was optimised to store the single photons, the success of this optimisation is evident as the recall efficiency for the single photons outperforms the recall of the coherent state at all times. From the relative scaling between these, it is evident that the single photon storage does not suffer from any additional degradation for extended storage times compared to the coherent state storage.  A detailed discussion on how the recall efficiency was calculated can be found in the Supplementary Materials.

Clearly GEM operates as well in the single photon regime as it does for weak coherent states when appropriately optimised\cite{Hosseini_Sparkes_Campbell_Lam_Buchler_2011} and we demonstrate a peak recall efficiency of 84\%, only one percent less than the demonstration by Wang \emph{et al.} \cite{Wang_Li_Zhang_Su_Zhou_Liao_Du_Yan_Zhu_2019}, but free of the complexity of vacuum systems and cold atom traps and we operate near continuously (only no storage during switching times). We note that during the switching phase of the field gradient indicated as the white background in Figures~\ref{expvscovssim}~\&~\ref{lifetime}, some additional signal is detected (not included in the recall efficiency). Possible explanations are noise from the change in currents in the coils or optical leakage of fields stored in the edges of the memory. See Supplementary material for more details.

 Performance of our source and memory system can be further improved: introducing a non-poled crystal into the SPDC cavity to utilise the clustering effect for intracavity filtering \cite{tsai_ultrabright_2018, bao_generation_2008, rielander_cavity_2016}, enhances the relevant single mode brightness of the source; optimising and improving the filtering systems to reduce the level of control field leakage; and optimising the memory geometry and magnetic field coil alignment, to improve coupling of the single photon wave function to the memory. All these steps will bring GEM even closer to its theoretical limit of 100\% recall efficiency.

\section*{Discussion}
Our demonstration highlights the advantages and viability of tailoring cavity-enhanced SPDC to produce memory-compatible single photons; together with the vacuum-free room temperature memory this provides a significant step towards a scalable quantum memory approach for networked quantum resources.
We present the first storage and recall of single photon states using the gradient echo memory protocol. Our achieved 84\% recall efficiency with a warm-vapour, vacuum-free, high duty-cycle, quantum memory is comparable with the best demonstrated quantum memory efficiency, $85\%$, which suffers intrinsically low duty-cycle, employing a high vacuum photon source and memory based on cold atoms. Our system is the first warm-vapor quantum memory to surpass the no-cloning limit with single photons. Through its always ready nature and absence of vacuum and cryo-technology our quantum memory system is suitable for integration with large scale networked quantum technology applications.

\section*{Materials and Methods}
\begin{figure*}[t]
    \centering
    \includegraphics[width=1\textwidth]{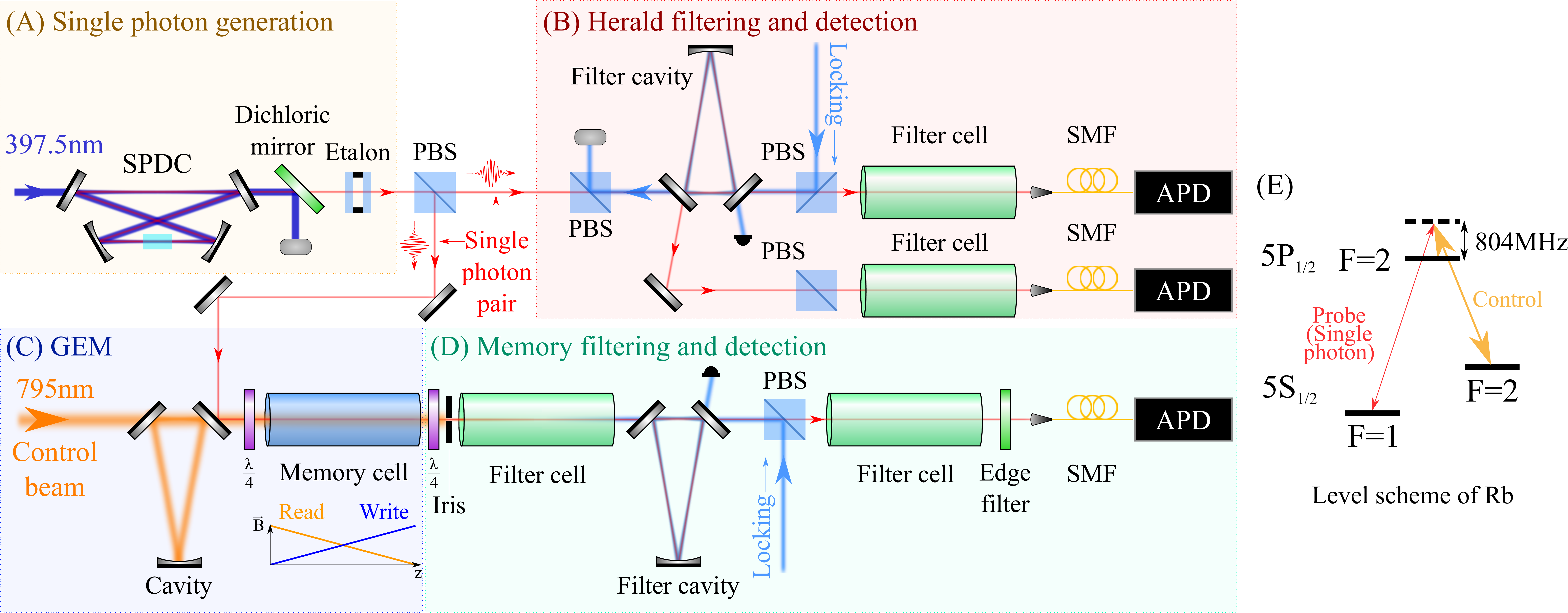}
    \caption{\textbf{Single photon storage setup.} \textbf{(A)} Single photon pairs are generated using cavity-enhanced spontaneous parametric down conversion.  A dichroic mirror filters out the 397.5~nm pump light and an etalon is used to suppress some of the unwanted single photon frequency modes.  The single photon pairs are separated using a polarizing beamsplitter (PBS) and fibre-coupled (not shown) with one sent for herald detection (B) and other sent to the quantum memory (C).  \textbf{(B)} The single photon is further filtered down to a single frequency mode with an optical cavity before being coupled to a single mode fibre (SMF) to be detected by a single photon avalanche photodiode (APD).  Both the cavity throughput and the reflected input are further filtered to remove scattering from the locking beam and detected to measure the single photon generation rate.  This rate is actively fed back to optimize the lock point of the SPDC cavity.  \textbf{(C)} The single photon and the control beam are combined on an output port of an optical cavity and sent into a hot vapor Rb cell.  A coil wrapped around the cell with varying pitch forms the magnetic field gradient to operate GEM.   \textbf{(D)} The control beam is mostly filtered using an iris, Rb filter cell, and optical cavity while the single photon is eventually detected using an APD. \textbf{(E)} Level scheme of Rb87 showing the utilised Raman transition for the GEM memory. The single photon together with the strong coherent control beam links the $5S_{1/2} F{=}1$ and $F{=}2$ ground states via off-resonant scattering of the $5P_{1/2} F=2$ state. During the storage phase the excitation is stored as a spinwave in the $5S_{1/2} F{=}2$ state. }
    \label{setup}
    \end{figure*}
\subsection*{Gradient Echo Memory}
In the warm, rubidium-vapour based Gradient Echo Memory (GEM) \cite{Hosseini_Sparkes_Campbell_Lam_Buchler_2011} memory scheme an optical lambda transition in rubidium 87 links the $F{=}1$ and $F{=}2$ states of the $5S_{1/2}$ ground state, see Figure~\ref{setup}E. The photon from the single-photon source addresses the probe transition (Rb87 D1 $5S_{1/2} F{=}1 \rightarrow 5P_{1/2} F{=}2$) and the bright ($50$mW) control beam (Rb87 D1 $5P_{1/2} F{=}2 \rightarrow 5S_{1/2} F{=}2$) transfers the excitation into the long lived ground state of the memory. Both optical fields are detuned by 804~MHz from resonance with the excited $5P_{1/2} F{=}2$ state.  The detailed laser setup to facilitate this can be found in the Supplementary Materials.  To gain control of the read and write times \cite{Hosseini_Sparkes_Hetet_Longdell_Lam_Buchler_2009} and avoid reabsorption of recalled signal in the memory, we apply a linear magnetic field gradient to the memory cell with coils of varying pitch.  This creates a spatially varying shift of the atomic resonance causing the memory to store different frequency components of the optical field at different spatial regions of the memory. Reversal of the magnetic field gradient and presence of the optical control beam initiate the recall of the stored signal.  As the GEM protocol eliminates unwanted reabsorption its theoretical efficiency can reach 100\%, and experimental results confirm that no noise is added to stored states making GEM highly suitable for single photon storage \cite{Hosseini_Campbell_Sparkes_Lam_Buchler_2011}.  To maximise our experimentally achievable memory efficiency we use a Rb87 enriched cell with 0.5 Torr of Kr buffer gas, heated to about $80^o$C with the coils and the cell contained inside a Mu-metal tube to shield from stray magnetic fields.

We keep the memory in write mode---magnetic field gradient and control field on---at all times until an idler photon detection event heralds a single photon storage event in the memory, see Figure~\ref{setup}B.  This triggers a set of relays to block the control field, ending the write phase, and flip the magnetic field gradients as shown in Figures~\ref{expvscovssim}A and \ref{expvscovssim}B. Once the gradient is fully reversed and the control field is turned back on, recall of the single photon commences. We switch the field gradient again after a predetermined time period depending on the storage time to reset the memory back into write mode (Figure~\ref{expvscovssim}A), ready to store for the next single photon from our source.

\subsection*{Single photon generation}
To match the strict spectral requirements of GEM, a cavity-enhanced SPDC source was designed to produce sub-MHz bandwidth biphoton pairs suitable for integration with the memory protocol \cite{Rambach_Nikolova_Weinhold_White_2016}.  A periodically poled potassium titanyl phosphate (ppKTP) crystal, cut for type-II quasi-phase matching, is used to produce orthogonally polarised photon pairs from a bow-tie cavity, as depicted in Figure~\ref{setup}A. Residual pump is removed with a dichroic mirror and ultra-narrow bandpass filter at 795~nm (FWHM 1.0~nm) after the source cavity. We use a polarising beam splitter to separate the idler and signal modes of the SPDC, which are fibre-coupled and sent to the heralding and memory arms respectively, as shown in Figure~\ref{setup}. 

When placed in a cavity, the otherwise broadband process of SPDC is modified so that emission into resonant modes are enhanced and, in the case where both downconverted modes are simultaneously resonant, further narrowed \cite{Herzog_theory-2008, slattery_2019}. Our source produces roughly 1000 frequency modes with a bandwidth of 429~kHz separated by the cavity's free spectral range (120.8~MHz) over the entire phase matching bandwidth of ${\approx}100$~GHz. GEM can only store one of these frequency modes at a time, thus a concatenation of external filtering systems shown in Figure~\ref{setup} is required to isolate the single central frequency mode relevant for GEM. The first stage of this system is an air-spaced flat surface etalon (SLS Optics, 75~GHz FSR, Finesse $\mathcal{F}{=}15$) placed immediately after the SPDC cavity before the idler and signal modes are separated. 

The herald is then further filtered by a mode-cleaning cavity ($764$~MHz FSR, $1$~MHz linewidth), locked with a counterpropagating beam resonant to a Rb85 transition and orthogonally polarized with respect to the herald photon (more on how this cavity was set up in the Supplementary Material).  The filtered herald photons are then passed through a Rb85 cell removing backscattered locking signal, before being detected by a fibre-coupled single photon counting module (SPCM). Similarly the stored signal photon is filtered after the memory with a different mode-cleaning cavity ($877$~MHz FSR, $1.6$~MHz linewidth) locked in a similar fashion. The convoluted transmission spectrum of these two mode-cleaning cavities, etalon and source cavity near perfectly isolate the central frequency mode of the SPDC (see Supplementary material, Figure~\ref{SingleFilterSpectrum}C), ensuring that coincident events detected between the herald and the stored photons will be of photons resonant with the memory transition.

To achieve a memory system ready to operate at 100\% duty-cycle, the photon source also must always be operational, without down-time due to load or re-locking sequences. The source cavity is locked to the pump beam using the Pound-Drever-Hall method \cite{drever_laser_1983, black_introduction_2001}, ensuring the SPDC process to be pumped at all times. Birefringence between the signal and idler modes is compensated with the half-waveplate `flip-trick' \cite{Rambach_Nikolova_Weinhold_White_2016} which reduces double resonance of both modes to one resonance condition. Fine temperature control of the SPDC crystal then achieves triple resonance by overlapping the resonance of the two down-converted modes at 795~nm (red) with that of the pump at 397.5~nm (blue). Slight temperature drifts allowed the resonances between the pump and down-converted photons to drift apart with time. To compensate we established a feedback-lock based on the discarded single photon counts of the herald's filter cavity, see Figure~\ref{setup}B. This signal comprises the majority of frequency modes and thus total number of produced single photons, enabling automated tracking of the triple resonance condition and continuous user-intervention free data collection for 8+ hours at a time. More details can be found in the Supplementary Material.
 
Efficient storage and recall requires a strong control beam (50~mW) to co-propagate with the single photons through the memory cell. Isolating the single photon signal after storage necessitates filtering of the control field as shown in Figure~\ref{setup}D. First an iris removed the part of the control beam that was larger than the signal beam. By tuning the control field to a Rb85 resonance (see Supplementary Material  Figure~\ref{filspec}) we gain 60dB of isolation from a Rb85 filter cell. However, atomic filter cells emit broadband fluorescence when absorbing resonant light \cite{heifetzSuperEfficientAbsorption2004}, which we removed with the aforementioned signal beam filter cavity. A second Rb85 filter cell suppresses leaking lock signal from this mode-cleaning cavity and adds further isolation against the control beam. Finally an edge filter removes the remaining fluorescence noise. The combined filtering chain measured 133dB suppression of the control beam.

\begin{figure}[tb]

      \includegraphics[width=0.9\linewidth]{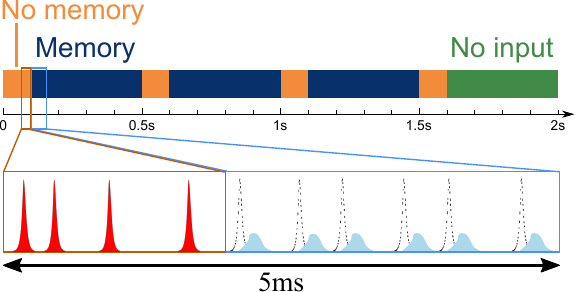}
    \captionof{figure}{\textbf{Data taking sequence.}  The data taking time sequence has a period of 2~s.  The memory is turned off for 0.1~s four times by turning the control field off (orange) measuring the rate of single photons injected into the memory.  The memory is operated for 1.2~s of the cycle (blue), and the final 0.4~s (green) operate with the control field on but with the incoming single photons blocked to measure the background noise.}  
    \label{datseq}

\end{figure}

\subsection*{Single photon detection}
Single photons are detected using avalanche photodiodes (Perkin Elmer SPCM-AQR-14-FC) and recorded using a time-tagging module (Roithner Lasertechnik, 100.1~ps time resolution). While both memory and source operated continuously, data taking was sequenced to compensate for long term drifts. The 2 second long sequence is broken into three different stages as shown in Figure~\ref{datseq}. In the first stage, labelled ``no memory", we measure the injected photon rate; in the second stage, labelled "memory", we store, then recall single photons; in the third stage, labelled "no input" we measure background counts, allowing for extraction of the single photon recall efficiency of the memory independent of long term drifts or variations in the experiment. In the memory stage, the recalled single photon signal is overlayed by leaked photons from the coherent control field contributing $\sim 5-10$k counts/s. Without detection in the coincidence basis between the signal and idler photon from the SPDC source recovery of the stored single photon signal would have been prohibited by such noise. Through our measurement sequencing we can subtract the noise using the scaled background data from the "no input" sequence, yielding the recall data shown in Figures \ref{expvscovssim} and \ref{lifetime} the data set without background subtraction is shown in the Supplementary Material as Figure \ref{noise} together with more details on the subtraction procedure. We emphasis that this noise is not caused by our memory protocol, GEM being a proven noiseless memory \cite{Hosseini_Campbell_Sparkes_Lam_Buchler_2011}, but leaked field that can be suppressed through improved spatial and spectral filtering.

\section*{Acknowledgments}
We thank Geoff T. Campbell and Daniel B. Higginbottom for valuable discussions during the initial stages of planning and prototyping the experiment.
\subsubsection*{Funding}
This research was supported by the Australian Research Council Centres of Excellence for Engineered Quantum Systems (EQUS, CE170100009) and Quantum Computation and Communication Technology (CE170100012).

\subsubsection*{Author contributions}
Building single photon source: WYSL, MR, TJW.  Building quantum memory: ACL.  Measurements and analysis: ACL and WYSL.  Combining source with memory: ACL, WYSL, ADT, KVP.  Supervision: BCB, PKL, AGW, TJW.  Writing - original draft: ACL, WYSL.  Writing - review and editing: ACL, WYSL, BCB, PKL, AGW, TJW.

\clearpage

\setcounter{figure}{0}
\renewcommand{\figurename}{Fig.}
\renewcommand{\thefigure}{S\arabic{figure}}

\section*{Supplementary Materials}
\subsection*{Laser Setup and Locking chains}
\begin{figure}[t]
    \centering
    \includegraphics[width=0.8\linewidth]{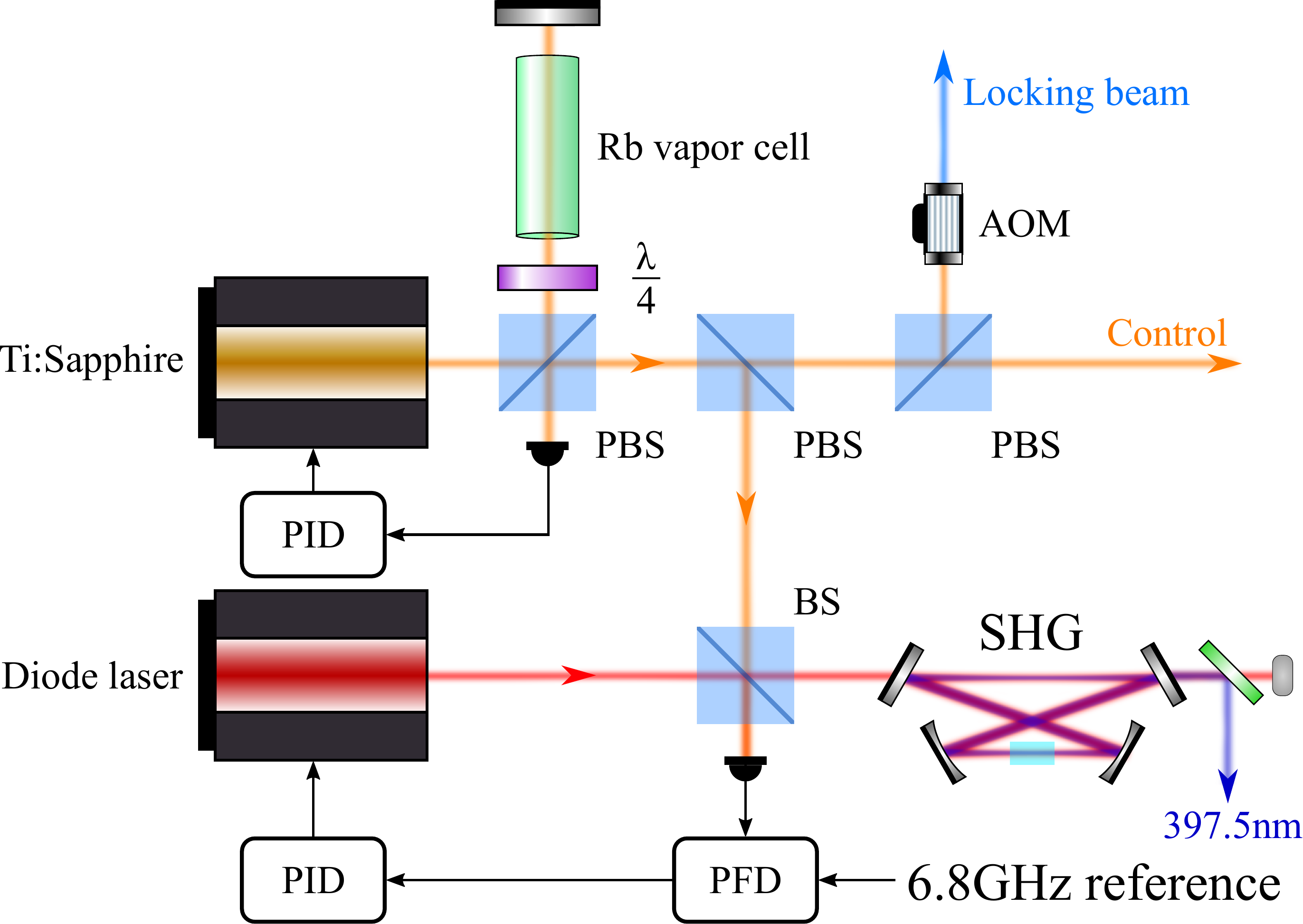}
    \caption{\textbf{Setup for generating all optical frequencies.}  The 795~nm Ti:Sapphire laser is frequency locked to a rubidium reference via saturated absorption.  The memory control field and cavity locking beams are derived from it and controlled with acoustic-optic modulators (AOM).  Some of the Ti:Sapphire light is combined on a 50:50 beamsplitter (BS) with light from a diode laser to form a beatnote that is detected with a phase frequency detector (PFD) and used to phase lock the diode laser 6.8~GHz away from the Ti:Sapphire.  The diode laser pumps a second harmonic generation (SHG) cavity which produces 397.5~nm light used to pump the SPDC cavity to generate single photons at the probe frequency.}
    \label{lasersetup}
\end{figure}

To operate GEM efficiently requires the frequencies of the probe light and the control field to match in their detuning from the $5P_{1/2} F{=}2$ state of Rb87.  By design, this detuning was chosen to have the control field resonant with the Rb85 D1 transition which allowed the use of filter cells that heavily attenuated control frequency light while having high transmission at the probe frequency.  The conceptual diagram of how all optical frequencies required were generated and referenced, including those for locking filter cavities is shown in Fig.~\ref{lasersetup}.  The 795~nm Ti:Sapphire laser is frequency locked to a rubidium reference cell on a Rb85 D1 transition which is 804~MHz blue detuned from the Rb87 D1 ($F{=}2 \rightarrow F{=}2$) control field transition shown in Fig.~\ref{setup}E.  We detect the beatnote of the Ti:Sapphire laser and the diode laser (MOGLabs 795 MSA), by mixing the beams of both lasers on a high speed photodiode. The signal is sent to a phase-frequency detector to phase lock the diode laser at a detuning of 6.8~GHz.  This stabilizes the diode laser 804~MHz away from the Rb87 D1 transition ($F{=}1 \rightarrow F{=}2$).  Using acoustic-optic modulators (AOMs) for finer frequency tuning, light from the Ti:Sapphire laser is also used to generate the two locking beams required to actively lock the filter cavities shown in Fig.~\ref{setup} of the main text.  The second harmonic generation (SHG) cavity is stabilised to the diode laser via Pound-Drever-Hall (PDH) locking \cite{drever_laser_1983,black_introduction_2001} for generating 397.5~nm light. This is then used to pump the cavity-enhanced SPDC process in Fig.~\ref{setup}A of the main text, which is in turn also locked via PDH locking to the pump light generated by the SHG cavity. Electro-optical modulators are used to create the sidebands for PDH locking. It is clearly essential for the GEM protocol that the central frequency mode of the SPDC cavity spectrum is correctly referenced to the desired detuned probe frequency of the Raman transition; this was achieved by referencing the pump light of the SHG cavity to the rubidium resonance via the beatnote, acting as an absolute reference for the down-conversion process.

\begin{figure*}[t]
    \centering
    \includegraphics[width=0.95\linewidth]{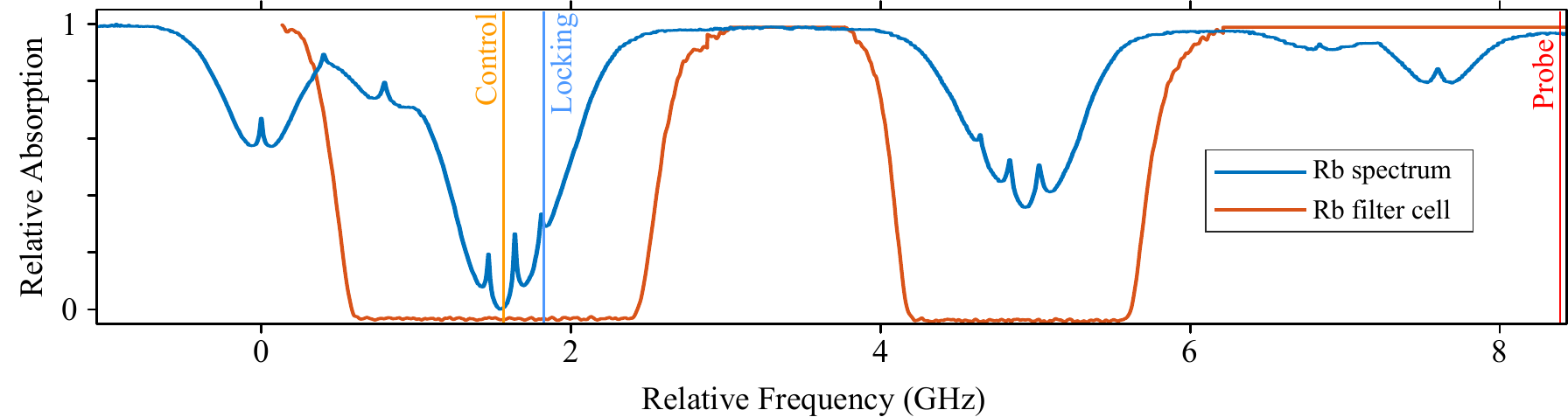}
    \caption{\textbf{Filter cell absorption spectra.}  Absorption of Rb filter cells and the frequency of the probe (single photons), control and locking beams}
    \label{filspec}
\end{figure*}

 In Fig.~\ref{filspec}, the different frequencies for the locking beams, control field for the memory and the probe frequency of the single photons are plotted relative to the spectrum of Rb85 and Rb87. Clearly the control field and locking beams fall within the broad absorption band of the hot rubidium atoms shown as the orange line, while the single photons at the probe beam frequency see a virtually transparent path in the absence of the lambda coupling facilitated by the control field. 

\subsection*{Monitoring the SPDC Cavity Lock Position}
Triply resonant parametric down conversion requires that the pump and both down converted fields are all simultaneously resonant in the enhancement cavity. As we are using type-II down-conversion, we have light fields of two wavelengths (397.5~nm, blue; and 795~nm, red) and with horizontal and vertical polarizations at 795~nm simultaneously present in the cavity. Ensuring triple resonance naively requires three free parameters. Using the flip-trick \cite{Rambach_Nikolova_Weinhold_White_2016}, we reduce the problem to only two free parameters. We use a piezo-mounted cavity mirror (stabilising with respect to the pump light), and the temperature of the down-conversion crystal, controlling the relative birefringence between the pump field and the down-converted modes. When the lock point on the blue resonance is exactly overlapped with the peak of the red resonance the generation rate of the photon pairs is maximised for a given pump power.  Environmental temperature drifts cause relative shifts between the red and blue resonances, hence shifting the set lock point away from the peak of the single photon resonance decreasing the generation rate. The cavity remains locked as the shifts are narrower than the PDH feature of the resonance used to lock the cavity. By design, the blue resonance ($\mathcal{F} \approx 8.5$) is much broader than the red ($\mathcal{F} \approx 181$) so that small adjustments to the offset of the PDH lock point can be made, allowing the lock point to follow the peak of the efficiency for generating down-converted photons.  
To do this we use the reflected single photons from the mode-cleaning cavity in the herald arm. This signal is proportional to the generation rate of the SPDC cavity as the number of modes rejected by the mode-cleaner is effectively constant and the most direct method of measuring the SPDC cavity drift alone. Any other available single photon signal (ie. at the end of the heralding arm, and after the memory) are subject to other cavity drifts masking the actual generation rate directly out of the SPDC cavity. To isolate these reflected photons from any leaked locking signal, another polarising beam splitter and filter cell is placed in this path, as depicted in Fig.~\ref{setup}B of the main paper.\\
As however we are effectively locking to the top of the peak, we implemented a dither lock that we dubbed PhD-lock (in honour of replacing the PhD student that used to perform this optimisation manually), that took a random step on the offset point of the PDC-cavity lock to try to increase the rejected single photon count rate. If count rate increased the PhD-lock would follow with a step in the same direction until the rate declined. Once a decline was detected the PhD-lock reversed polarity and stepped in the opposite direction until the count rate again declined leading to an effective dither around the maximal count rate. Our PhD-lock used a dynamic step size, where each adjustment was proportional to the difference in single photon counts averaged over \textit{n} clock cycles from the previous \textit{n} cycles. Changing the number of integration cycles allowed trading the PhD-lock's robustness for speed and vice-versa.
On occasion the drift of the optimal lock position exceeded the capture range of the PDH lock, which could then be regained by a small temperature adjustment of the crystal oven to re-overlap them again. The stability of the entire lock chain was such that continuous unsupervised data acquisition over approximately half a day was possible.

\subsection*{Single Photon Filtering Spectrum}

\begin{figure}[t]
    \centering
        \includegraphics[width=1\linewidth]{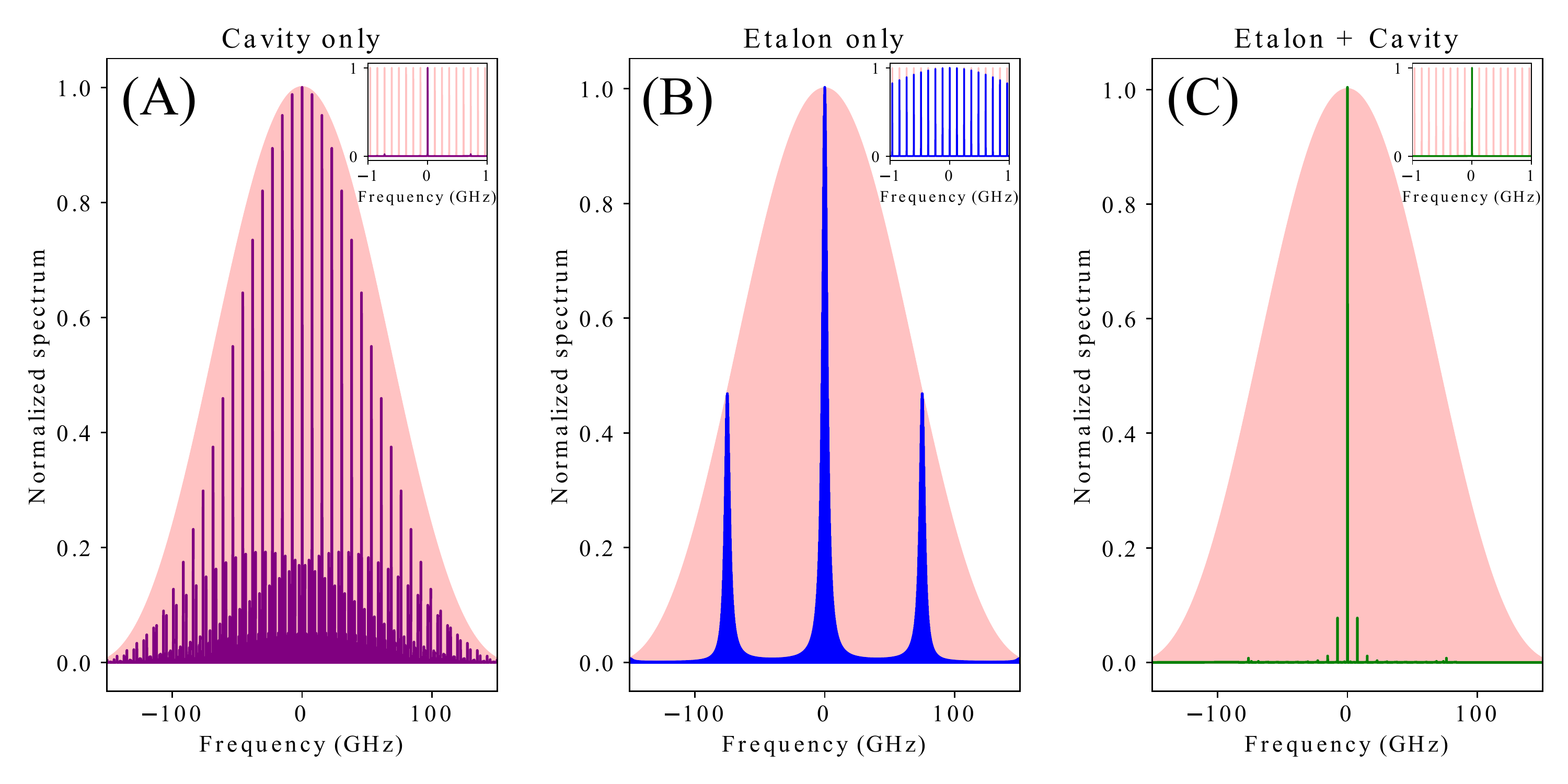}
        \caption{\textbf{Cavity only, etalon only and combined cavity and etalon filtering of SPDC.}  The simulated spectrum for the single photons after filtering with just \textbf{(A)} the herald arm mode-cleaner cavity, \textbf{(B)} the etalon after the SPDC cavity or \textbf{(C)} combined cavity and etalon filtering.  The pink shading is the original unfiltered output spectrum of the SPDC.  The insets in each are the zoomed in spectrum centered on the central frequency mode.}
    \label{SingleFilterSpectrum}
\end{figure}

The biphoton pairs produced by the cavity-enhanced SPDC source exist as a frequency comb spanning approximately 100~GHz with a 120.8~MHz free spectral range and a linewidth of $429 \pm 10$~kHz for each mode. The  spectrum is formed as the convolution of the crystal's phase matching bandwidth, and the cavity's spectrum.  To reduce the frequency comb to a single frequency mode, external filtering was required. We combined an etalon (SLS Optics, FSR $= 75$~GHz, $\mathcal{F} = 15$) and a mode-cleaning cavity (FSR = $761 \pm 14$~MHz, high finesse mode with a linewidth of $1.1 \pm 0.4$~MHz). A simulation of the resulting spectra of each filtering component is presented in Fig.~\ref{SingleFilterSpectrum}(A and B), and the final spectrum from the concatenated chain in Fig.~\ref{SingleFilterSpectrum}C. The most conservative measure of the power in this central frequency mode is $76\%$. For memory recalled events we are detecting the events in coincidence. So while the herald has undergone the filtering as described above, the signal photon also passed through the same etalon, but was filtered by different mode-cleaning cavity in the memory arm (with a FSR $= 881$~MHz). The mismatch in FSR between the two mode-cleaning cavities thus ensures that all coincident events result from only the central frequency mode as all other modes are suppressed in either the heralds or the memory arm. 
Note that while the etalon spectrum in Fig.~\ref{SingleFilterSpectrum}B contain 3 main transmission sections, within each section the frequency modes separated by the FSR of the SPDC cavity still exists, as seen in the inset of the figure. 

\subsection*{Filtering cavity setup}
Both filtering cavities in the herald and memory arms were set up the same way with a locking beam generated by an AOM from the Ti:Sapphire combined with a PBS to lock the cavities in the backward direction.  The locking beam had an orthogonal polarization to the single photons and locked the cavities in the low finesse mode while the single photons were being filtered in the high finesse mode.  To optimize the throughput of the single photons, the locking beam frequency was fine-tuned to maximize the single photon count rate measured by the single photon detectors.  Backscattering of the locking beam added to the noise counts measured by the single photon detectors so a Rb85 filter cell was placed after the cavity which absorbed any backscattered light.  These cavities, in addition to the etalon, was sufficient for filtering the SPDC output to a single frequency mode.  The mode-cleaning cavity after the memory also served for control filtering; it is used to suppress both control frequency light that passed through the first filter cell and also any wide-band fluorescence caused by the filter cell absorbing the control beam.  The second filter cell after the cavity was also used to provide further suppression of control light.  Finally, a long pass wavelength filter (cutoff at 790~nm) removed the rest of the fluorescence before the single photons were coupled into a single mode fibre for spatial filtering and sent to a single photon detector.

\subsection*{Quantifying memory efficiency}
\begin{figure}
    \centering
    \includegraphics[width=0.8\linewidth]{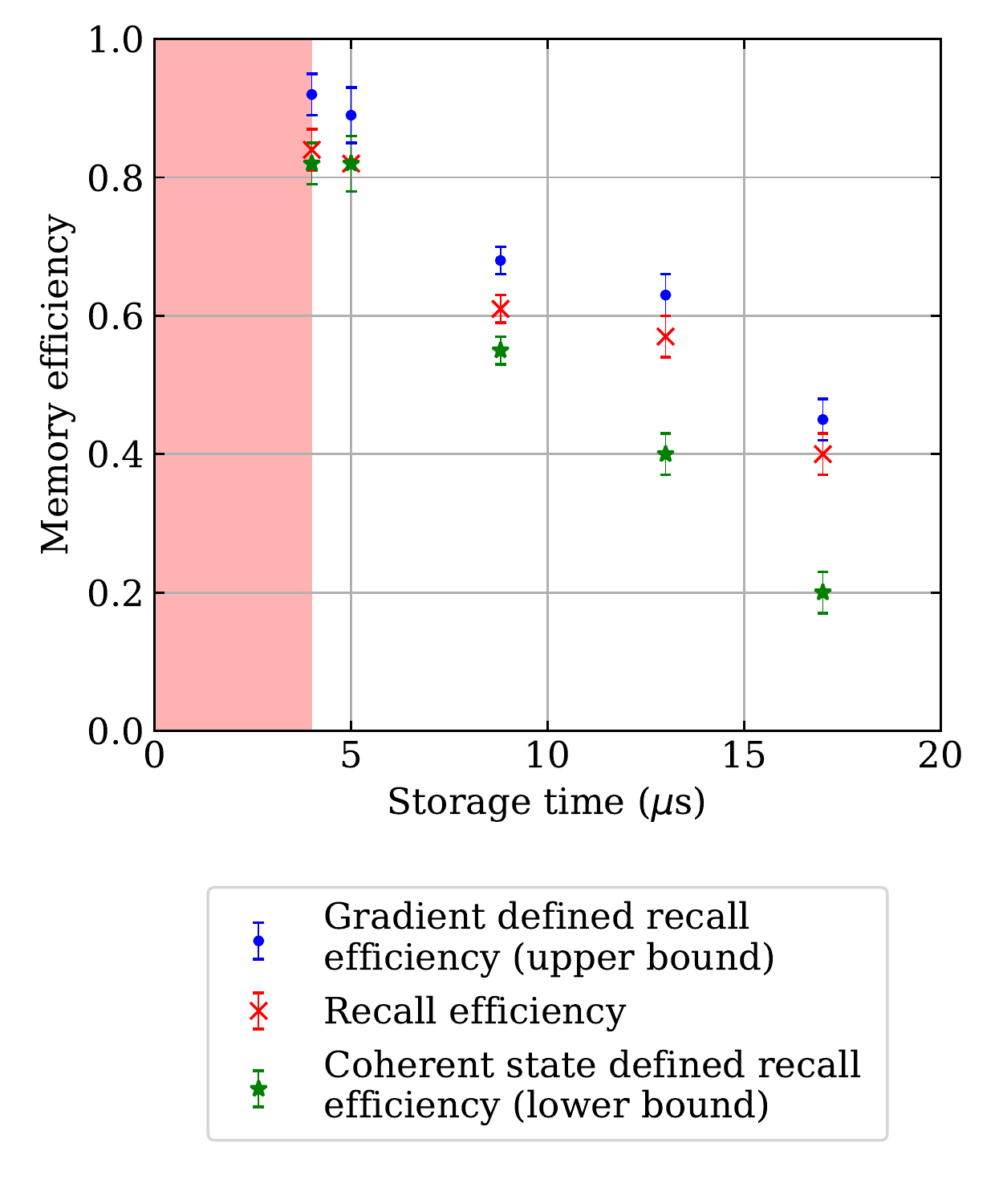}
    \caption{\textbf{Memory recall efficiency at different storage times.}  The red shaded area identifies storage times shorter than our minimum storage duration.  We show here the results of our upper bound, the reported recall efficency and the lower bound. As upper bound serves the gradient defined efficiency (blue dots), which considers all photons from the memory while the magnetic field gradient is negative (read mode).  The reported recall efficiency (red crosses) considers photons when the gradient is reversed \emph{and} the control field is on.  Finally, the lower bound is set by the coincidences integrated over the times at which the coherent state recall was measured (green stars).}
    \label{effvstime_all}
\end{figure}

The recall efficiencies measured from GEM depend directly on the integration time over which the counts were considered. The upper bound to the efficiency is found by starting the integration at the time when the gradient of the applied magnetic field is reversed, thus switching the memory from write mode to read mode. This was calculated as a check to ensure that the total amount of recalled and transmitted light did not exceed what was initially stored. The gradient flip is the required condition for a recall, but the recall is not initiated until also the control beam is applied. The value for the phase where both of these conditions are met, is thus the reported efficiency in the main text. A lower bound was determined based on the time window during which the recalled coherent state was measured.  As noted earlier, there were key differences in the temporal shape between the single photons and coherent state, thus the coherent state defined time window does not extend to all times when recalled single photons are expected; the efficiencies calculated this way serves as a lower bound to the recall efficiency.  These upper and lower bounds to recall efficiency  are shown together with the recall efficiency in Figure \ref{effvstime_all}.  All three forms of efficiency were generated with the scaled coincidences as explained in the single photon detection section of the main text.

\subsection*{Leakage during magnetic field switching}
The small level of recall observed coming from the memory in the storage phase during which the memory's magnetic fields are switching could potentially have been due to: a combination of the arrival of single photons from other frequency modes in the SPDC acting as control photons completing a Raman or EIT transition inducing some recall; and the memory being transparent to signal photons that have arrived during the magnetic field switching time due to the sub-unity heralding efficiency.  Another possible explanation for this observation is electronic noise in the detection system induced by the magnetic fields switching.  Further investigation is required to fully understand this.  The method in which this data was collected and processed means that the final results, like those shown in Fig.~\ref{expvscovssim} of the main text, show \textit{only} the effect of sending single photons to the quantum memory.  This rules out the possible effect of additional photons caused via four wave mixing due to the control field, as well as other noise caused by the input single photons as they are too weak and far detuned from any atomic transition to cause similar effects. Such conclusions are in agreement to what was demonstrated in \cite{Hosseini_Campbell_Sparkes_Lam_Buchler_2011} which showed the GEM scheme in warm vapor does not introduce additional noise to its recall. 

\subsection*{Background subtraction methodology for memory data}

\begin{figure}[t]
    \centering
    \includegraphics[width=0.95\linewidth]{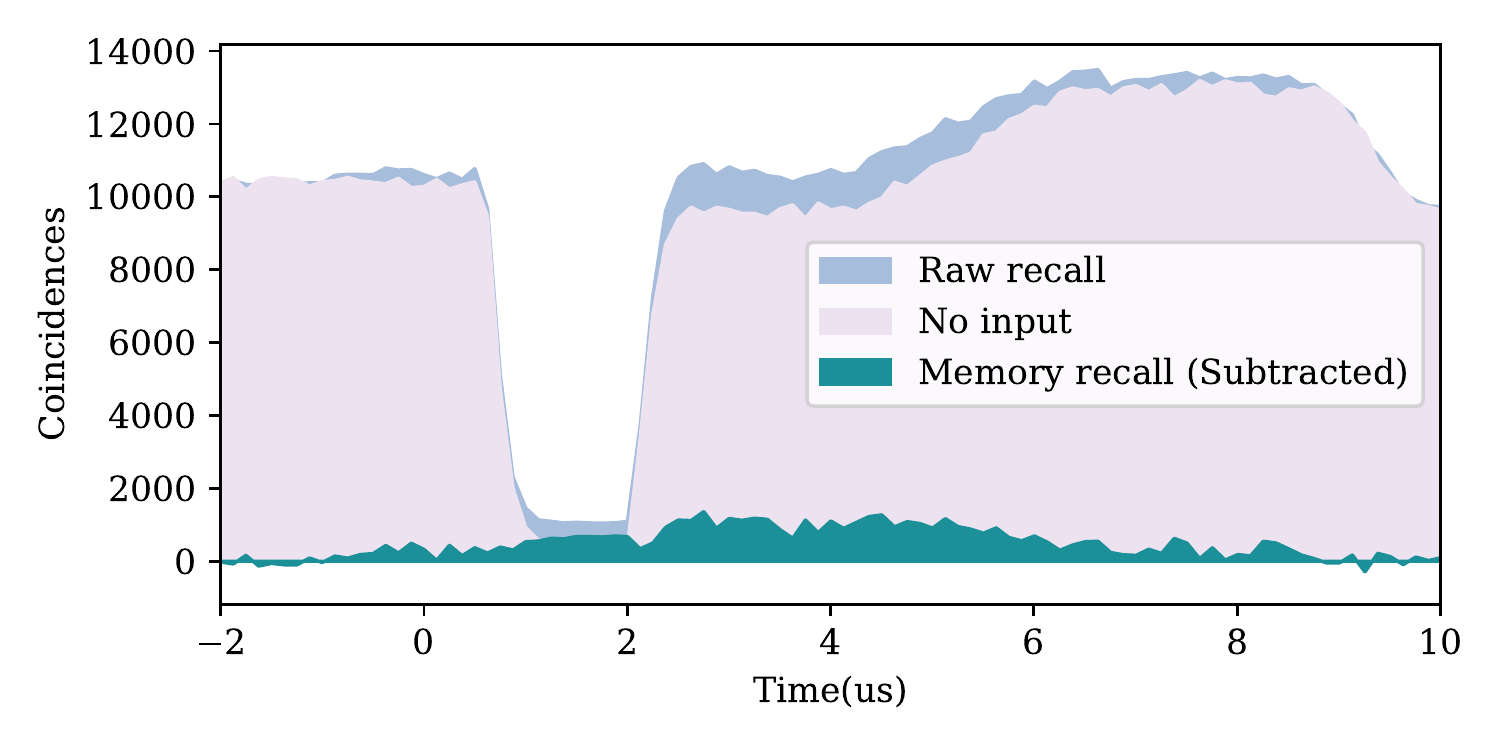}
    \vspace{-0.4cm}
    \caption{\textbf{Example data set for the coincidence collection for 4~$\mu$s storage.} The purple area are coincidences collected during the ``No input'' data acquisition phase. The blue area labeled Raw recall is data collected during the ``Memory'' phase and the green area is the difference between the ``Memory'' and ``No input'' data sets that was then used to provide the recalled signal.}
    \label{noise}
    \vspace{-0.90cm}
\end{figure}
While our filtering chains used for the control field and the locking light beams for the filter cavities provided excellent suppression of these fields on the single photon counter for the signal single photons, some small amount of light from these sources still reached this detector. The total filtering chain provided 133~dB isolation as mentioned in the main text, but despite this significant reduction, we still observed 5-10k counts/s as noise background. As our scheme attributes any single photons detected during the recall phase before the magnetic field gradient is flipped back as a coincidence detection with the herald that triggered the recall procedure, we need to remove these noise events from the real recalls of single photons. Figure \ref{noise} shows the collected data for the 4~$\mu$s storage from the ``Memory'' and ``No input'' phases of the data collection cycle.  During the ``No memory'' phase used to collect the input single photon signal shown in Fig. \ref{expvscovssim}C (of the main paper) such subtraction was not required as no control field was present. 
The success of this method is evident from our data presented in the main paper. As the control field leakage amount drifted with alignment and temperature changes in the filter cells, this motivated the sequenced data acquisition method ensuring that ``Memory'' and ``No input'' data was taken at effectively equal conditions. 


\begin{thebibliography}{10}

\bibitem{kimble_internet}
H.~J. Kimble, {\it Nature\/} {\bf 453}, 1023 (2008).


\bibitem{Wehner2018}
S.~Wehner, D.~Elkouss, R.~Hanson, {\it Science\/} {\bf 362}, eaam9288 (2018).

\bibitem{Wei2022}
S.-H. Wei, {\it et~al.\/}, {\it Laser \& Photonics Reviews\/} p. 2100219
  (2022).

\bibitem{Gottesman2012}
D.~Gottesman, T.~Jennewein, S.~Croke, {\it Physical Review Letters\/} {\bf 109}
  (2012).

\bibitem{Jiang2007}
L.~Jiang, J.~M. Taylor, A.~S. S{\o}rensen, M.~D. Lukin, {\it Physical Review A
  - Atomic, Molecular, and Optical Physics\/} {\bf 76} (2007).

\bibitem{Nickerson2014}
N.~H. Nickerson, J.~F. Fitzsimons, S.~C. Benjamin, {\it Physical Review X\/}
  {\bf 4} (2014).

\bibitem{Komar2014}
P.~K{\'{o}}m{\'{a}}r, {\it et~al.\/}, {\it Nature Physics\/} {\bf 10}, 582
  (2014).

\bibitem{Lee2019}
S.~W. Lee, T.~C. Ralph, H.~Jeong, {\it Phys. Rev. A\/} {\bf 100}, 052303
  (2019).

\bibitem{Heshami2016}
K.~Heshami, {\it et~al.\/}, {\it Journal of Modern Optics\/} {\bf 63}, 2005
  (2016). PMID: 27695198.

\bibitem{Briegel1998}
H.~J. Briegel, W.~D{\"{u}}r, J.~I. Cirac, P.~Zoller, {\it Phys. Rev. Lett.\/}
  {\bf 81}, 5932 (1998).

\bibitem{Heinze_2013_PRL}
G.~Heinze, C.~Hubrich, T.~Halfmann, {\it Phys. Rev. Lett.\/} {\bf 111}, 033601
  (2013).

\bibitem{Zhong_2015_Nat}
M.~Zhong, {\it et~al.\/}, {\it Nature\/} {\bf 517}, 177 (2015).

\bibitem{Hosseini_Sparkes_Campbell_Lam_Buchler_2011}
M.~Hosseini, B.~Sparkes, G.~Campbell, P.~Lam, B.~Buchler, {\it Nature
  Communications\/} {\bf 2} (2011).

\bibitem{Hsiao_2018_PRL}
Y.-F. Hsiao, {\it et~al.\/}, {\it Phys. Rev. Lett.\/} {\bf 120}, 183602 (2018).

\bibitem{Cho_Campbell_Everett_Bernu_Higginbottom_Cao_Geng_Robins_Lam_Buchler_2016}
Y.-W. Cho, {\it et~al.\/}, {\it Optica\/} {\bf 3}, 100 (2016).

\bibitem{England_PRL_2015}
D.~G. England, {\it et~al.\/}, {\it Phys. Rev. Lett.\/} {\bf 114}, 053602
  (2015).

\bibitem{Guo_NatComm_2018}
J.~Guo, {\it et~al.\/}, {\it Nature Communications\/} {\bf 10}, 148 (2019).

\bibitem{Knill_Laflamme_Milburn_2001}
E.~Knill, R.~Laflamme, G.~J. Milburn, {\it Nature\/} {\bf 409}, 46–52 (2001).

\bibitem{Fan_CrossKerrBreakdown_2013}
B.~Fan, {\it et~al.\/}, {\it Phys. Rev. Lett.\/} {\bf 110}, 053601 (2013).

\bibitem{Vural-18}
H.~Vural, {\it et~al.\/}, {\it Optica\/} {\bf 5}, 367 (2018).

\bibitem{Tsai_2020_PRR}
P.-J. Tsai, Y.-F. Hsiao, Y.-C. Chen, {\it Phys. Rev. Research\/} {\bf 2},
  033155 (2020).

\bibitem{Heller_2021} L.~Heller, J.~Lowinski, K.~Theophilo, Padr\'on-Brito, and H.~de~Riedmatten, arXiv:2111.08598 (2021).

\bibitem{Yu_2021} Y.~Yu, {\it et~al.\/}, 
{\it Phys. Rev. Lett.\/} {\bf 127}, 160502 (2021).

\bibitem{Wang_Li_Zhang_Su_Zhou_Liao_Du_Yan_Zhu_2019}
Y.~Wang, {\it et~al.\/}, {\it Nature Photonics\/} {\bf 13}, 346-351  (2019).

\bibitem{Cao_Hoffet_Qiu_Sheremet_Laurat_2020}
M. Cao, F. Hoffet, S. Qiu, A.~S. Sheremet, J. Laurat, {\it Optica.\/} {\bf 7}, 1440 (2020).

\bibitem{SFWMNoise}
K.~Heshami, {\it et~al.\/}, {\it Journal of Modern Optics\/} {\bf 63}, 2005
  (2016).

\bibitem{Michelberger_2015}
P.~S. Michelberger, {\it et~al.\/}, {\it New Journal of Physics\/} {\bf 17},
  043006 (2015).

\bibitem{ORCA-PhysRevA.97.042316}
K.~T. Kaczmarek, {\it et~al.\/}, {\it Phys. Rev. A\/} {\bf 97}, 042316 (2018).

\bibitem{Hosseini_Sparkes_Hetet_Longdell_Lam_Buchler_2009}
M.~Hosseini, {\it et~al.\/}, {\it Nature\/} {\bf 461}, 241–245 (2009).

\bibitem{Hosseini_Campbell_Sparkes_Lam_Buchler_2011}
M.~Hosseini, G.~Campbell, B.~M. Sparkes, P.~K. Lam, B.~C. Buchler, {\it Nature
  Physics\/} {\bf 7}, 794–798 (2011).

\bibitem{Rambach_Nikolova_Weinhold_White_2016}
M.~Rambach, A.~Nikolova, T.~J. Weinhold, A.~G. White, {\it APL Photonics\/}
  {\bf 1}, 096101 (2016).

\bibitem{Wootters1982}
W.~K. Wootters, W.~H. Zurek, {\it Nature\/} {\bf 299}, 802 (1982).

\bibitem{Hong}
C.~K. Hong, L. Mandel, {\it Phys. Rev. Lett.\/} {\bf 56}, 58 (1986).

\bibitem{Pittman} 
T.~B. Pittman, B.~C. Jacobs, J.~D. Franson,  {\it Optics Communications\/} {\bf 246}, 545 (2005).

\bibitem{first-g2}
H.~J. Kimble, M.  Dagenais, L. Mandel, {\it Phys. Rev. Lett.\/} {\bf 39}, 691 (1977).


\bibitem{tsai_ultrabright_2018}
P.-J. Tsai, Y.-C. Chen, {\it Quantum Science and Technology\/} {\bf 3}, 034005
  (2018).

\bibitem{bao_generation_2008}
X.-H. Bao, {\it et~al.\/}, {\it Phys. Rev. Lett.\/} {\bf 101}, 190501
  (2008).

\bibitem{rielander_cavity_2016}
D.~Rieländer, A.~Lenhard, M.~Mazzera, H.~d. Riedmatten, {\it New Journal of
  Physics\/} {\bf 18}, 123013 (2016).

\bibitem{Herzog_theory-2008}
U.~Herzog, M.~Scholz, O.~Benson, {\it Phys. Rev. A\/} {\bf 77}, 023826 (2008).

\bibitem{slattery_2019}
O.~Slattery, L.~Ma, K.~Zong, X.~Tang, {\it Journal of Research of the National
  Institute of Standards and Technology\/} {\bf 124}, 124019 (2019).

\bibitem{drever_laser_1983}
R.~W.~P. Drever, {\it et~al.\/}, {\it Appl. Phys. B\/} {\bf 31}, 97 (1983).

\bibitem{black_introduction_2001}
E.~D. Black, {\it American Journal of Physics\/} {\bf 69}, 79 (2001).

\bibitem{heifetzSuperEfficientAbsorption2004}
A.~Heifetz, {\it et~al.\/}, {\it Optics Communications\/} {\bf 232}, 289–293
  (2004).

\end{thebibliography}
\end{document}